\documentclass[preprint,number,times,sort&compress]{elsarticle}

\usepackage{amsmath}
\usepackage{siunitx}
\usepackage{graphicx}
\usepackage{color}
\usepackage[active]{srcltx} 

\newcommand{\const}{\mathrm{const}}
\newcommand{\mx}{{\mathrm{max}}}
\newcommand{\pc}{{\mathrm{pc}}}

\newcommand{\erf}{{\mathrm{erf} \,}}

\newcommand{\ex}{{\mathrm{exc}}}
\newcommand{\bs}{{\mathrm{base}}}
\newcommand{\av}{{\mathrm{av}}}

\newcommand{\al}{\alpha}

\newcommand{\ga}{\gamma}
\newcommand{\de}{\delta}

\newcommand{\la}{\lambda}

\newcommand{\sig}{\sigma}
\newcommand{\De}{\Delta}

\usepackage{amsthm}

\theoremstyle{remark}

\begin{document}

\title{Modeling of a heat capacity peak and an enthalpy jump for a paraffin-based phase-change material}

\author[nr,pha]{Igor Medved'\corref{cor1}}
\ead{igor.medved@fsv.cvut.cz}
\author[nr,pha]{Anton Trn\'{\i}k}
\ead{atrnik@ukf.sk}
\author[nr]{Libor Voz\'{a}r}
\ead{lvozar@ukf.sk}

\cortext[cor1]{Corresponding author. Tel.: +421 37 6408 616; fax:
+421 37 6408 556.}

\address[nr]{Department of Physics, Constantine the Philosopher
University, 94974 Nitra, Slovakia}
\address[pha]{Department of Materials Engineering and Chemistry, Czech Technical University, 16629 Prague, Czech Republic}

\begin{abstract}
Rubitherm RT~$27$ is a paraffin-based phase-change material (PCM) in which a change between a solid and liquid phase is used to store/release thermal energy. Its enthalpy and heat capacity, as measured in a quasistatic regime by adiabatic scanning calorimetry, has a single distinct jump and peak, respectively, at about $\SI{27.3}{\celsius}$. We present a microscopic development from which the jump and peak can be accurately fitted and that could be analogously applied even to other PCMs. It enables us to determine the baseline and excess part of the heat capacity and thus the latent heat associated with the phase change. It is shown to be about $84 \%$ of the total enthalpy change that occurs within $\SI{5}{\celsius}$ from the peak maximum position. The development is based on the observation that PCMs often have polycrystalline structure, being composed of many single-crystalline grains. The enthalpy and heat capacity measured in experiments are therefore interpreted as superpositions of many contributions that come from the individual grains.
\end{abstract}


\maketitle


\section{Introduction}

When materials absorb or release heat, their temperature varies in general. However, if a phase change occurs in materials, then the temperature only slightly varies, even though a large amount of energy is stored or released. Only after the phase change is over does the temperature begin to rise or fall significantly. Therefore, materials with a phase change, or phase-change materials (PCMs), are of great interest in the applications where there is demand for thermal energy storage with a high density (within a small temperature range) and/or where a temperature level needs to be maintained. Examples are solar energy storage \cite{Ken07}, space heating and cooling of buildings \cite{Cab11,Kuz11,Soa13}, cold storage applications \cite{Cab12}, data storage applications \cite{Wutt11}, and industrial applications in textiles and clothing systems \cite{SO12}.

It is well known that during a phase change between two phases (such as melting/freezing between a solid and liquid phase) the enthalpy vs.~temperature plot shows a sudden jump, while the heat capacity vs.~temperature plot shows a distinct peak. The presence of such rounded jumps and peaks is attributed mostly to non-equilibrium effects; if heat exchange were carried out quasistatically and the studied sample were macroscopically \emph{large}, the jumps and peaks would become infinitely sharp. In some experiments, however, heat capacity peaks keep their finite width even at rather slow heating rates \cite{Red09,Sch09}. For example, when adiabatic scanning calorimetry (ASC) is applied, very slow scanning rates can be achieved (down to $\SI{0.5}{mK.min^{-1}}$) so that thermodynamic equilibrium of the investigated samples is ensured \cite{Glor11a}. This suggests that finite jumps and peaks need not be a pure non-equilibrium phenomenon, but it should be possible to obtain them even within an equilibrium approach.

In this paper we wish to present such an approach and demonstrate that it can predict rounded jumps/peaks from experiments with very good accuracy. The approach is based on the observation that the crystalline state of PCMs has usually a polycrystalline structure, being composed of many single-crystalline grains some of which have just few tens of nanometers in diameter \cite{Wutt11}. We thus propose to interpret an experimentally measured jump/peak as a superposition of many contributions coming from the individual grains (see Section~\ref{sec: poly}). Due to finite-size effects, the jumps/peaks from the \emph{small} grains are sharp, yet of finite width. In addition, they are mutually shifted. Therefore, when they are superimposed, the so obtained result can fit experimental data with very good precision (see Section~\ref{sec: appl}).

The starting point of our approach is a microscopic theory \cite{BoKo95} from which enthalpy jumps and heat capacity peaks in a single grain can be obtained (see Section~\ref{sec: single}). It should be noted that lately there has been a number of studies of PCMs using various microscopic techniques, such as molecular dynamics simulations \cite{Par09,Par10,Par10a,Ell11,Ell12,Ell13,Ell14,Anan14}, density-functional calculations \cite{Jon08,Jon09,Jon09a,Jon11,Jon14}, a cellular automata approach \cite{Wri08}, classical nucleation theory simulations \cite{Bur12}, and a statistical theory of crystallization \cite{Pet13}. Most of these works focus on specific materials (one or more of the alloys Ge$_2$Sb$_2$Te$_5$, Sb$_2$Te$_3$, GeTe, AgInSbTe, and Ga-Sb) due to their practical importance in digital memory technologies.

\begin{figure}
  \centering
  \includegraphics[width=\linewidth]{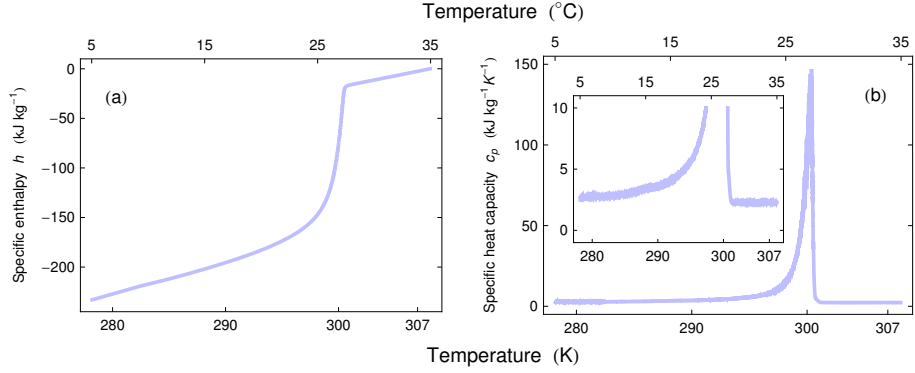}\\
\caption{(a) The specific enthalpy and (b) the specific heat capacity of Rubitherm RT~$27$ obtained from ASC measurements performed in \cite{Glor11}. Either thick line represents more than $21$ thousand data points. The inset in (b) shows the heat capacity near its two foot regions.}
  \label{fig: data}
\end{figure}

As a specific material, we shall consider a paraffin-based PCM called Rubitherm RT~$27$ in which a change between a solid and liquid phase is used to store/release thermal energy in various civil engineering applications. Its enthalpy and heat capacity were measured in \cite{Glor11} using adiabatic scanning calorimetry. Thus, it should be plausible to apply a quasistatic approach to describe these results in which
the enthalpy has a single distinct jump and the heat capacity has a single distinct peak. The phase-change temperature was determined to be $\SI{27.3}{\celsius}$ for heating and $\SI{27.2}{\celsius}$ for cooling. We shall focus on the heating part of the temperature dependences (the corresponding experimental data are shown in Fig.~\ref{fig: data} and listed in Table~\ref{tab: RT data}), because, on closer inspection, they are more representative than those for the cooling run.
\begin{table}
\footnotesize
\centering
\begin{tabular}{lllll}
  \hline \hline
  Quantity & Symbol & Value & Unit & Reference
\\
  \hline
  Phase change temperature & $T_\pc$ & $300.5$ & K & \cite{Glor11}
\\
  Enthalpy change between $T_\pc \pm \SI{5}{K}$& $\De h$ & $165$ & $\si{kJ.kg^{-1}}$ & \cite{Glor11}
\\
  Solid phase density & $\rho_s$ & $880$ & $\si{kg.m^{-3}}$ & Producer
\\
  Liquid phase density & $\rho_l$ & $760$ & $\si{kg.m^{-3}}$ & Producer
\\
  \hline \hline
\end{tabular}
\caption{The basic properties of Rubitherm RT~$27$ (produced by Rubitherm GmbH, Germany).} \label{tab: RT data}
\end{table}

\section{Single-crystalline PCMs: Extremely sharp peaks and jumps}
\label{sec: single}

We shall consider a phase change that occurs between two phases. Then a jump in the specific enthalpy, $h$, is expected to interpolate between the specific enthalpies, $h_1$ and $h_2$, of the two phases; i.e.,
\begin{subequations} \label{eq: h,c single}
\begin{equation} \label{eq: h single}
  h = h_1 + (h_2 - h_1) \, \eta,
\end{equation}
where the quantity $0 < \eta < 1$ describes the precise form of the interpolation. Since $\eta = (h-h_1)/(h_2-h_1)$, it has the meaning of a normalized enthalpy and is dimensionless. It is further expected that a peak in the heat capacity, $c_p$, is the sum of the excess and baseline heat capacities,
\begin{equation}
  c_p = c_\ex + c_\bs.
\end{equation}
Similarly to $h$, the baseline capacity should interpolate between the heat capacities, $c_1$ and $c_2$, of the two phases,
\begin{equation}
  c_\bs = c_1 + (c_2 - c_1) \, \eta',
\end{equation}
where $0 < \eta' < 1$ is the corresponding interpolation function. On the other hand, the excess capacity has the shape of a peak, for it is associated with the phase change itself. It may be written as the product
\begin{equation}
  c_\ex = c_0 \ga,
\end{equation}
\end{subequations}
where $c_0$ is its maximal value (about two orders of magnitude larger than the single-phase capacities $c_1$ and $c_2$), and $0 < \ga \leq 1$ is a dimensionless quantity describing the peak in $c_\ex$.

At present there is no universal microscopic theory of phase changes for realistic models of materials that would predict jumps in the enthalpy and peaks in the heat capacity as given in Eqs.~\eqref{eq: h,c single}. Nevertheless, for simplified models, called lattice gases, such a general theory was already developed \cite{BoKo90,BoKo95}. It is appropriate only for processes in which temperature changes are performed quasistatically. Moreover, since lattice gases are suitable for the description of changes between crystalline phases, a PCM that we can thus describe must have a perfect, single-crystal microstructure. This is plausible for a solid phase of the studied PCM, but it is somewhat approximative for a liquid phase.

If we invoke the theory from \cite{BoKo95}, then the results from Eqs.~\eqref{eq: h,c single} can be indeed obtained. Namely, it follows that the two interpolating functions are identical, $\eta \approx \eta'$, and can be both approximated by the function $J(x) = (1+\tanh x)/2$, while the peak function $\ga$ can be approximated by the function $P(x) = \cosh^{-2} x$. The functions $J$ and $P$ are similar in shape to the Gaussian error function and bell curve, respectively, but they approach their limiting values at a slower, exponential rate. The shorthand $x$ and the maximal value $c_0$ are given as
\begin{equation} \label{eq: xT, c0, DT0}
  x = 2 \, \frac{T - T_\mx}{\De T_0},
  \quad
  \De T_0 = \frac{4 k_B T_\pc^2}{\ell m},
  \qquad \qquad
  c_0 = \frac \ell{\De T_0} = \frac{\ell^2 m}{4 k_B T_\pc^2},
\end{equation}
where $T_\pc$ is the temperature of the phase change, $\ell = h_2(T_\pc) - h_1(T_\pc)$ is the specific latent heat associated with the change, $m$ is the sample mass (assumed to be constant), and $k_B$ is the Boltzmann constant. Note that $c_0$ and $\ell$ is equal to the \emph{height} and \emph{area} of the excess heat capacity peak, respectively, and $\De T_0 = \ell / c_0$ corresponds to its \emph{half-width}. The temperature $T_\mx$ is the maximum position of the total heat capacity $c_p$. It is slightly shifted with respect to the phase-change temperature due to surface effects (the influence of the surroundings),
\begin{equation} \label{eq: Tm}
  T_\mx - T_\pc \approx \frac{\sig S}{\ell m} \; T_\pc.
\end{equation}
Here $\sig$ is the difference between the specific (per unit area) surface free energies of the two phases and $S$ is the surface size of the sample, both evaluated at $T = T_\pc$.
%

\begin{figure}
  \centering
  \includegraphics[width=\linewidth]{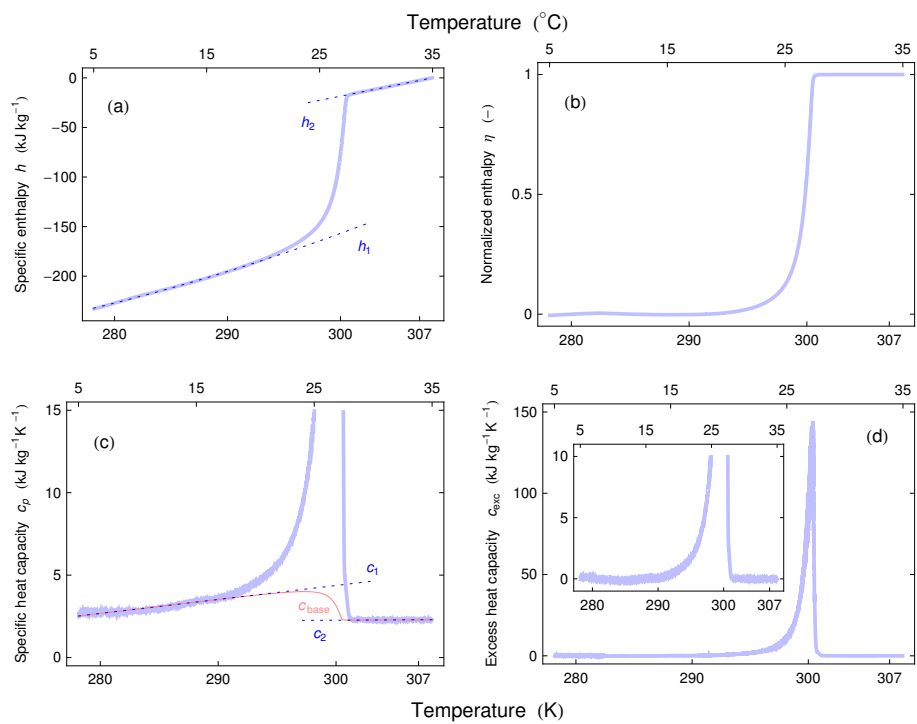}\\
\caption{(a) The single-phase enthalpies $h_1$ and $h_2$ (the dotted lines) determined for Rubitherm RT~$27$ by fitting the data from Fig.~\ref{fig: data}(a) by quadratic polynomials. (b) The corresponding normalized enthalpy $\eta$. (c) The single-phase capacities $c_1$ and $c_2$ (the dotted lines) determined for Rubitherm RT~$27$ by fitting the data from Fig.~\ref{fig: data}(b) by linear polynomials, and the corresponding baseline heat capacity $c_\bs$ calculated for $\eta$ from part (b). (d) The excess heat capacity obtained as the difference $c_p - c_\bs$. The inset shows its two foot regions in greater detail.}
  \label{fig: data norm}
\end{figure}

Using the experimental data from Fig.~\ref{fig: data} for Rubitherm RT~$27$, we may use quadratic fits to determine the enthalpies $h_1$ and $h_2$ and linear fits to determine the heat capacities $c_1$ and $c_2$, and then calculate the normalized enthalpy $\eta$, baseline heat capacity $c_\bs$, and excess heat capacity $c_\ex$ (see Fig.~\ref{fig: data norm}).
The latter has the height $\SI{144.4}{kJ.kg^{-1}.K^{-1}}$, half-width $\SI{0.64}{K}$, and area $\SI{138.5}{kJ.kg^{-1}}$. In the above theoretical results these should coincide with $c_0$, $\De T_0$, and $\ell$, respectively. This might be perhaps true for samples of just few nanometers in size, such as for nano-encapsulated PCMs \cite{Peng13}. However, for samples of few micrometers in diameter, Eq.~\eqref{eq: xT, c0, DT0} with $T_\pc$ and $\rho$ from Table~\ref{tab: RT data} predicts a peak that is about \emph{eight orders of magnitude sharper and taller} than the one observed experimentally (if the latent heat is kept unchanged). In fact, the same conclusion follows for any PCM for which the heat capacity peak has the height $c_0$, half-width $\De T_0$, and area (latent heat) $\ell$ of orders $\SI{100}{kJ.kg^{-1}.K^{-1}}$, $\SI{1}{K}$, and  $\SI{100}{kJ.kg^{-1}}$, respectively.

Therefore, the theoretical description based on Eqs.~\eqref{eq: h,c single} and \eqref{eq: xT, c0, DT0} cannot be used to accurately reproduce various experimental data, and a more sophisticated approach must be adopted.


\section{Polycrystalline PCMs: Wide peaks and jumps}
\label{sec: poly}

The main reason why Eqs.~\eqref{eq: h,c single} and \eqref{eq: xT, c0, DT0} yield results that may be inconsistent with experiment is the assumption that a PCM has a perfect, single-crystalline microstructure. If we consider a PCM that is polycrystalline, consisting of a number of single-crystal grains, then we will be able to fit experimental data with theoretical results with very good precision. A single-crystalline PCM is a special case when there is just one grain.

\subsection{Model of polycrystalline PCMs}

The grains, $G$, may be of various sizes and their surroundings may affect them in different ways. For simplicity, we will assume that the grains are of spherical shape and mutually independent (non-interacting) and that possible effects of void spaces between the grains are neglected. Then the enthalpy and heat capacity of a PCM sample is the sum of the enthalpies and heat capacities coming from its individual grains. The specific enthalpy and capacity may thus be expressed as the weighted averages, $h = \sum_G w_G h_G$ and $c_p = \sum_G w_G c_G$, of the grain specific enthalpies, $h_G$, and capacities, $c_G$, respectively. The weight of a given grain $w_G = m_G/m$ is equal to the fraction of its mass in the sample.

Applying Eqs.~\eqref{eq: h,c single} -- \eqref{eq: Tm} to $h_G$ and $c_G$ (with the sample mass $m$ and sample surface $S$ replaced by the grain mass, $m_G$, and grain surface, $S_G$, respectively), we get
\begin{subequations} \label{eq: h,c poly}
\begin{equation} \label{eq: h poly}
  h \approx h_1 + (h_2 - h_1) \, J_\av,
  \qquad
  c_p = c_\ex + c_\bs
\end{equation}
with
\begin{equation}
  c_\ex \approx c_0 \, P_\av,
  \qquad
  c_\bs \approx c_1 + (c_2 - c_1) \, J_\av.
\end{equation}
\end{subequations}
These results have the same form as for single-crystalline samples:
\begin{itemize}
\item[(a)]
the specific enthalpy interpolates between the single-phase specific enthalpies $h_1$ and $h_2$;
\item[(b)]
the specific heat capacity is the sum of the excess and baseline heat capacities;
\item[(c)]
the excess heat capacity is the product of $c_0$ and a dimensionless peak function;
\item[(d)]
the baseline capacity interpolates between the single-phase specific heat capacities $c_1$ and $c_2$, similarly to the enthalpy.
\end{itemize}
This time, however, the interpolation is described by the average $J_\av = \sum_G w_G J(x_G)$ of the grain jump functions $J(x_G)$. In addition, the capacity peak is described by the average $P_\av = \sum_G w_G^2 P(x_G)$ of the products $w_G P(x_G)$ of the weights and gain peak functions $P(x_G)$, because $c_\ex = \sum_G w_G [ c_{0G} P(x_G) ]$ with $c_{0G} = \ell^2 m_G / 4 k_B T_\pc^2 = c_0 (m_G/m) = c_0 w_G$. In the special case of a single-crystalline PCM, there is only one grain $G$ with $w_G = 1$, and Eq.~\eqref{eq: h,c poly} reduces back to Eqs.~\eqref{eq: h,c single}.

We anticipate that Eq.~\eqref{eq: h,c poly} can predict \emph{much wider and smaller} heat capacity peaks and much wider enthalpy jumps for polycrystalline PCMs than for single-crystalline PCMs. Indeed, if a PCM is composed of many grains, then the jump and peak functions $J(x_G)$ and $P(x_G)$ from different grains are mutually shifted and of various widths, depending on the grain sizes and surface effects. Therefore, when multiplied by the (usually very small) terms $w_G$ and $w_G^2$, respectively, and summed together, the resulting averages $J_\av$ and $P_\av$ could be much wider and, in the latter case, much smaller. Moreover, the positions of $J(x_G)$ and $P(x_G)$ for different grains are inversely proportional to the grain diameter ($T_\mx - T_\pc \propto S_G/m_G \propto 1/d$). Hence, $J(x_G)$ and $P(x_G)$ are unevenly distributed in a given temperature range, so that their averages $J_\av$ and $P_\av$ and, therefore, the enthalpy jumps and heat capacity peaks are expected to be \emph{asymmetric} in general, in agreement with experimental data.

In the following it is sufficient to focus on the excess heat capacity $c_\ex$, because the baseline heat capacity $c_\bs$ and enthalpy $h$ are obtained from the average $J_\av$ by Eq.~\eqref{eq: h,c poly}, and the latter can be calculated from $c_\ex$ by integration,
\begin{equation} \label{eq: cJ from c_exc}
  J_\av \approx \frac1{\ell} \int_0^T c_\ex (T) \, dT,
\end{equation}
as can be easily verified.

To evaluate the excess heat capacity $c_\ex$, we shall rewrite it in a more convenient form, using the PCM density, $\rho$, and grain diameter, $d$, both evaluated at the phase-change temperature $T_\pc$. We express the grain mass and surface as $m_G = \rho  V_G = \pi \rho d^3 / 6$ and $S_G = \pi d^2$, respectively. Then
%
%
the variations in the grain heat capacities $c_G$ between various grains are only due to the grain diameter $d$ and its surface free energy difference $\sig$ (see Eqs.~\eqref{eq: xT, c0, DT0} and \eqref{eq: Tm}). So, if we classify the grains according to their values of $d$ and $\sig$, we may express the excess heat capacity as a double sum,
\begin{equation} \label{eq: c_exc double av}
  c_\ex = c_0 \sum_{i=1}^n N_i \, \Big( \frac{d_i}D \Big)^6 P_i,
  \qquad
  P_i = \sum_{j=1}^{N_i} \nu_{ij} \, P (x_{ij}),
\end{equation}
where $D$ is a diameter of the PCM sample at $T_\pc$. The first sum is over all grain diameters $d_1, \dots, d_n$. The second sum is over all values $\sig_1, \dots, \sig_{N_i}$ of the surface free energy differences in the grains of a fixed diameter $d_i$; the number of these grains is denoted as $N_i$. The quantity $\nu_{ij}$ is the fraction of the grains of diameter $d_i$ whose value of the surface free energy difference is $\sig = \sig_j$. The shorthand $x_{ij}$ stands for $x$ evaluated for a grain with a diameter $d_i$ and $\sig = \sig_j$.

Since the numbers $N_i$ and weights $\nu_{ij}$ are unknown for Rubitherm RT~$27$, we shall consider simple forms of these weights to obtain an explicit formula for $c_\ex$ from Eq.~\eqref{eq: c_exc double av}.

\subsection{Surface effects}

We will assume that the grains are created in a random process so that the boundary conditions for various grains are irregular, which is why $\sig$ changes from one grain to another. We let $\sig_0$ denote the mean value of $\sig$. In addition, since we consider $\sig$ to be random and related to the grain boundary, we shall assume that its fluctuation (standard deviation) is inversely proportional to the square root of the number $M_i$ of atoms lying on the grain boundary, $\De\sig_i \propto 1/\sqrt{M_i}$. Thus, $\De\sig_i = b_0/d_i$, where $b_0 > 0$ is a constant.
Taking the values $\sig_1, \dots, \sig_{N_i}$ to be equally spread, for simplicity, we may approximate $\nu_{ij}$ for a grain of diameter $d_i$ by the Gaussian form,
\begin{equation} \label{eq: la Gauss}
  \nu_{ij} \approx \la_i (\sig_j) \, \de\sig_i,
  \qquad
  \la_i (\sig) = \frac1{\sqrt{2\pi} \, \De\sig_i} \,
  e^{- \frac12 \, \big( \frac{\sig - \sig_0}{\De\sig_i} \big)^2},
\end{equation}
where $\de\sig_i = (\sig_{N_i} - \sig_1)/(N_i-1)$ is the distance between two adjacent values $\sig_j$.

\begin{figure}
  \centering
  \includegraphics[width=\linewidth]{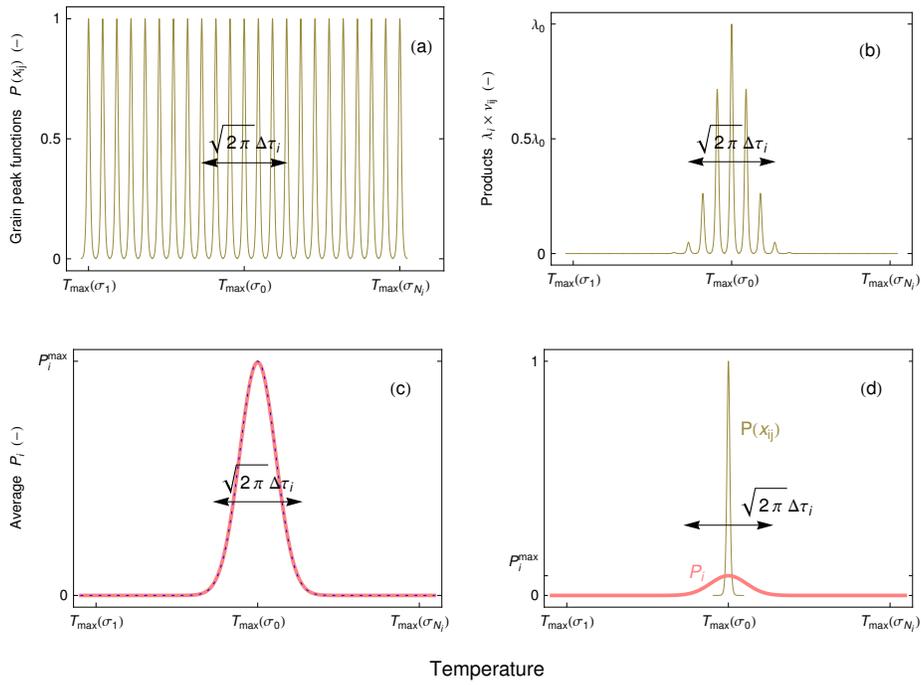}\\
\caption{(a) The grain peak functions $P(x_{ij})$ for $N_i = 200$ (every ninth peak is shown). The value $\sqrt{2\pi} \; \De\tau_i$ is $11.9$ times larger than the half-width $\De T_i$ of all grain peaks. (b) The same grain peaks multiplied by the weights $\nu_{ij}$ from Eq.~\eqref{eq: la Gauss}. A maximal value $\la_0$ of the products is indicated. (c) The average $P_i$ of the grain peaks obtained numerically (the full line) and from the approximation from Eq.~\eqref{eq: Pi} (the dotted line). The maximal value (height) is $P_i^\mx = \De T_i / \sqrt{2\pi} \; \De\tau_i$. (d) The average $P_i$ is smaller and wider than a grain peak function by the factor $P_i^\mx$.}
  \label{fig: aver}
\end{figure}

According to Eqs.~\eqref{eq: xT, c0, DT0} and \eqref{eq: Tm}, the half-width and maximum positions of the grain peaks $P(x_{ij})$ in the average $P_i$ from Eq.~\eqref{eq: c_exc double av} are given by
\begin{equation}
  \De T_i = \frac{24 k_B T_\pc^2}{\pi \ell \rho d_i^3},
  \qquad
  T_\mx^{ij} = \Big( 1 + \frac{6\sig_j}{\ell \rho d_i} \Big) T_\pc.
\end{equation}
While the half-width is the same for all peaks $P(x_{ij})$ in $P_i$, their maximum positions vary proportionally to $\sig_j$. Thus, the peaks with different $\sig_j$ are spread over a range between the temperatures $T_\mx^{i1}$ and $T_\mx^{iN_i}$ corresponding to the maximal and minimal value $\sig_1$ and $\sig_{N_i}$, respectively. The dominant peaks $P(x_{ij})$ in the average $P_i$ are, however, those with a high weight $\nu_{ij}$; i.e., the peaks corresponding to $\sig_j$ between $\sig_0 - \De\sig_i$ and $\sig_0 + \De\sig_i$. These are spread over a narrower range of half-width
\begin{equation}
  \De\tau_i = \frac{6 \De\sig_i}{\ell \rho d_i} \, T_\pc.
\end{equation}
The average $P_i$ strongly depends on the ratio of the half-widths $\De T_i$ and $\De\tau_i$. Indeed, if $\De\tau_i$ is much smaller than $\De T_i$, there are only small shifts between the grain peaks $P(x_{ij})$, and the average $P_i$ is practically the same as a peak function for a single grain of diameter $d_i$. On the other hand, if $\De\tau_i$ is much larger than $\De T_i$, the grain peaks $P(x_{ij})$ are spread over a wide temperature range, and the average $P_i$ is \emph{much wider and smaller} than a grain peak (see Fig.~\ref{fig: aver}). Namely \cite{MPH15},
\begin{equation} \label{eq: Pi}
  P_i \approx \frac{\De T_i}{\sqrt{2\pi} \; \De\tau_i} \, e^{- y_i^2/2 },
  \qquad
  y_i = \frac{T - T_\mx^i}{\De\tau_i}
      = \frac{d_i}{b_0} \,
      \Big( \frac{\ell \rho d_i}6 \, \frac{T-T_\pc}{T_\pc} - \sig_0 \Big),
\end{equation}
provided the ratio
\begin{equation} \label{eq: cond}
  \frac{\sqrt{2\pi} \; \De\tau_i}{\De T_i}
  = \Big( \frac\pi2 \Big)^{3/2} \, \frac{b_0 d_i}{k_B T_\pc}
  \gg 1.
\end{equation}
Here $T_\mx^i$ is the maximal temperature $T_\mx$ for a grain of diameter $d_i$ taken at the mean value $\sig = \sig_0$. Thus, while every grain peak $P(x_{ij})$ has the height $1$ and half-width $\De T_i$, the average $P_i$ has, according to Eq.~\eqref{eq: Pi}, the height $\De T_i / \sqrt{2\pi} \; \De\tau_i \ll 1$ and half-width about $\sqrt{2\pi} \; \De\tau_i \gg \De T_i$.
%
%

\subsection{Peak in the excess heat capacity: the final formula }

To get the excess heat capacity, it remains to perform the averaging over the grain diameters $d_1, \dots, d_n$. The simplest case is that the diameters are equally spread and that there is an equal number of grains of a given diameter,
\begin{equation}
  N_i = \const = \frac{D^3}{d_0^3},
  \qquad d_0^3 = \sum_{i=1}^n d_i^3,
\end{equation}
where we used that $\sum_i N_i (\pi d_i^3/6)$ must be equal to the total PCM volume $\pi D^3/6$. Then Eqs.~\eqref{eq: c_exc double av} and \eqref{eq: Pi} yield
\begin{equation} \label{eq: c_exc explicit}
  c_\ex \approx \frac{\ell^2 \rho}{6\sqrt{2\pi} \; b_0 d_0^3 T_\pc} \,  \sum_{i=1}^n d_i^5 \, e^{- y_i^2/2}.
\end{equation}
Thus, the excess heat capacity is a sum of peaks whose maxima are located at $T_\mx^i  \propto 1/d_i$. Thus, as $d_i$ increases, these positions get closer and closer to the phase-change temperature $T_\pc$, but their mutual distances are not equal. Consequently, $c_\ex$ is a sum of unevenly distributed peaks and will in general be \emph{asymmetric}. This is a pure finite-size effect. The special case when $c_\ex$ is symmetric can occur only if all peaks have the same position, $\sig_0 = 0$.

\section{Results and discussion}
\label{sec: appl}

\begin{figure}
  \centering
  \includegraphics[width=\linewidth]{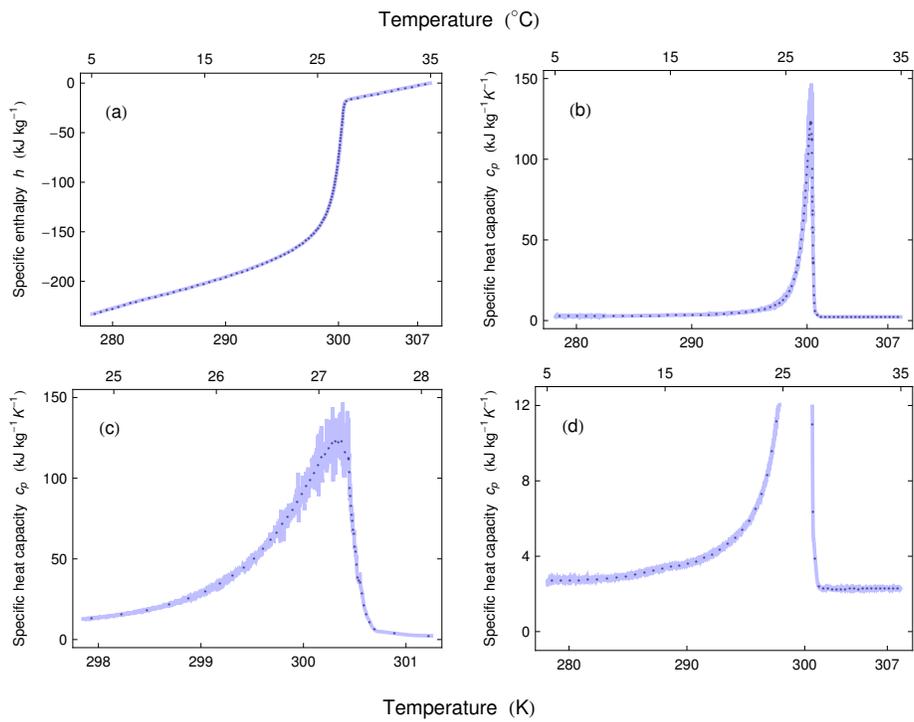}\\
\caption{The original data from Fig.~\ref{fig: data} (lines) and the averaged data (dots) obtained for (a) the specific enthalpy and (b) the specific heat capacity of Rubitherm RT~$27$. In (c) and (d) the averaged data for the specific heat capacity near the peak maximum and near the two peak foots, respectively, are shown in detail.}
  \label{fig: data av}
\end{figure}

\begin{figure}
  \centering
  \includegraphics[width=\linewidth]{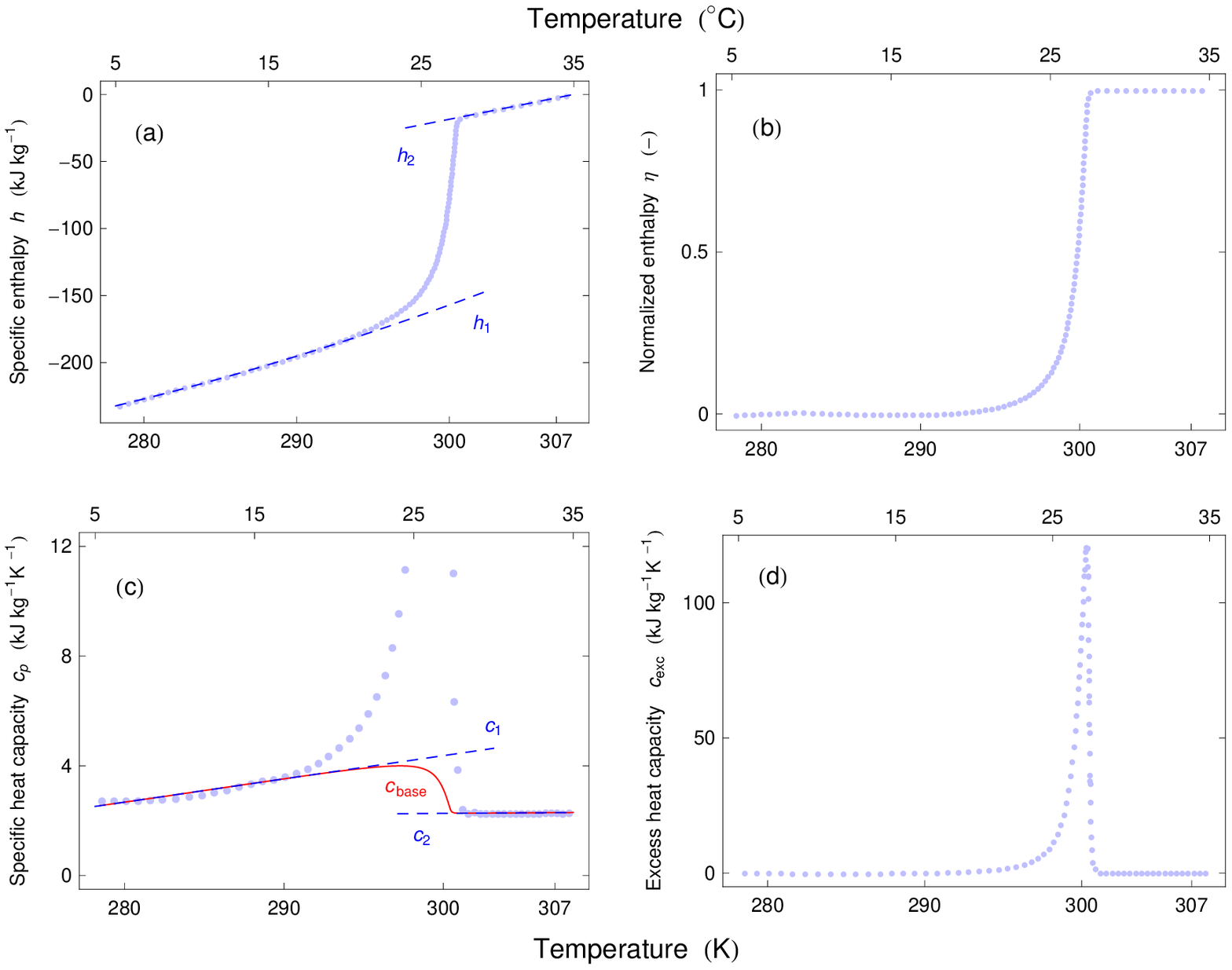}\\
\caption{(a) The single-phase enthalpies $h_1$ and $h_2$ (the dashed lines) determined for Rubitherm RT~$27$ by fitting the averaged data from Fig.~\ref{fig: data av}(a) by quadratic polynomials. (b) The corresponding normalized enthalpy $\eta$. (c) The single-phase capacities $c_1$ and $c_2$ (the dashed lines) determined for Rubitherm RT~$27$ by fitting the averaged data from Fig.~\ref{fig: data av}(b) by linear polynomials, and the corresponding baseline heat capacity $c_\bs$ calculated for the interpolating function $\eta$ from part (b). (d) The excess heat capacity obtained as the difference $c_p - c_\bs$.}
  \label{fig: data norm av}
\end{figure}

Let us apply the above theoretical results to fit the experimental data for Rubitherm RT~$27$ plotted in Fig.~\ref{fig: data}. Since the data for the heat capacity are quite oscillating, especially near its maximum, let us consider also their averaged version that is much smoother and should be more representative (see Figs.~\ref{fig: data av} and \ref{fig: data norm av}). In the fitting procedure presented below we choose the minimal grain diameter $d_1 = \SI{10}{nm}$ and number of different grain sizes $n = 300$. The maximal grain diameter will be allowed to attain a range of values, $d_n = \SI{0.1}{\micro m}$, $\SI{0.15}{\micro m}$, $\dots$, $\SI{0.5}{\micro m}$, to observe the sensitivity of the results to this parameter. In addition, the sample density at $T_\pc$ will be estimated as an average of the solid and liquid densities, $\rho = (\rho_s + \rho_l)/2 = \SI{820}{kg.m^{-3}}$ (see Table~\ref{tab: RT data}). Thus, there are four parameters in Eq.~\eqref{eq: c_exc explicit} for $c_\ex$ that remain to be fitted to the data: the phase-change temperature $T_\pc$, specific latent heat $\ell$, mean value $\sig_0$, and width $b_0$. To determine them, four independent properties of $c_\ex$ taken from the experimental data must be fitted by theoretical expressions. We shall proceed as follows.

First, we consider the \emph{area} under the peak exhibited by $c_\ex$. Since $c_\ex$ is an average of the grain peaks all of which have the area equal to $\ell$ (see Section~\ref{sec: single}), the peak of $c_\ex$ has also the area equal to $\ell$. Calculating the peak area for the original data in Fig.~\ref{fig: data norm}(d) and averaged data in Fig.~\ref{fig: data norm av}(d), we get
\begin{equation}
  \ell = \SI{138.5}{kJ.kg^{-1}},
  \qquad
  \ell = \SI{139.2}{kJ.kg^{-1}},
\end{equation}
respectively, which is about $84$ \% of the total enthalpy change in the range between $\SI{295.5}{K}$ and $\SI{305.5}{K}$ (see Table~\ref{tab: RT data}). The reason of this discrepancy is that the excess heat capacity decreases to zero in both foot regions, and so it has a smaller area than the total heat capacity.

\begin{figure}
  \centering
  \includegraphics[width=\linewidth]{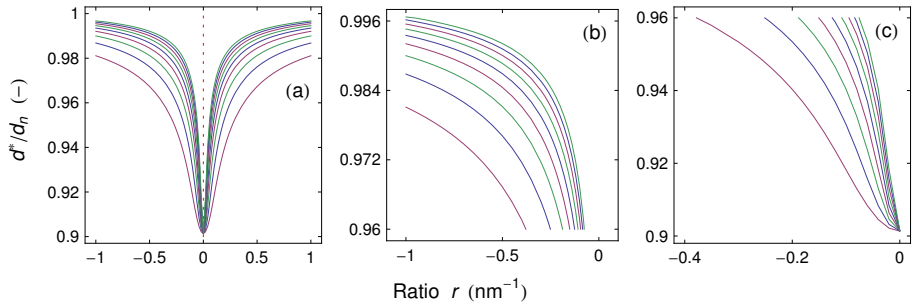}\\
\caption{(a) The size $d^*$ (relative to $d_n$) specifying the maximum position $T^*$ of the excess heat capacity in dependence on the ratio $r = \sig_0 / b_0$ calculated for $d_n = \SI{0.1}{\micro m}$, $\SI{0.15}{\micro m}$, $\dots$, $\SI{0.5}{\micro m}$ (the lines from bottom to top). In (b) and (c) the size $d^*$ for negative ratios $\sig_0 / b_0$ and its the maximal and minimal values, respectively, are shown in detail.}
  \label{fig: dst}
\end{figure}

Second, we consider the \emph{maximum position}, $T^*$, of the peak exhibited by $c_\ex$. If we express $T^*$ in a form similar to $T_\mx$,
\begin{equation} \label{eq: T*}
  T^* = \Big( 1 + \frac{6\sig_0}{\ell \rho d^*} \Big) \, T_\pc,
\end{equation}
where $d^*$ is a suitable diameter, then the condition $d c_\ex (T^*)/dT = 0$ for the maximum may be rewritten as
\begin{equation} \label{eq: L* eq}
  \sum_{i=1}^n d_i^8 (d_i - d^*) \, e^{- z_i^2/2} = 0,
  \qquad
  z_i = y_i (T^*) = \frac{\sig_0 d_i}{b_0} \, \Big( \frac{d_i}{d^*} - 1 \Big).
\end{equation}
The diameter $d^*$ is the solution to this equation. It depends only on the ratio $r = \sig_0 / b_0$ (and not on particular experimental data). This dependence can be calculated numerically and is plotted in Fig.~\ref{fig: dst}.

\begin{figure}
  \centering
  \includegraphics[width=\linewidth]{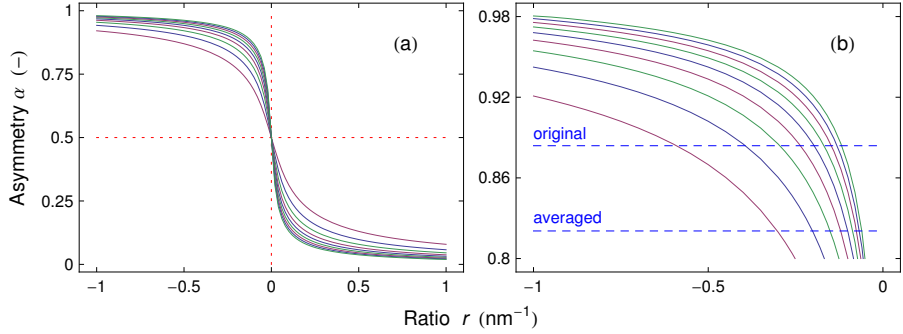}\\
\caption{(a) The asymmetry factor $\al$ in dependence on the ratio $r = \sig_0 / b_0$ calculated for the size $d^*$ from Fig.~\ref{fig: dst}. (b) The asymmetry factor $\al$ for negative ratios $\sig_0 / b_0$ and near its maximal values. The experimental values of $\al$ for the original and averaged data are shown as dashed lines.}
  \label{fig: al}
\end{figure}

\begin{figure}
  \centering
  \includegraphics[width=\linewidth]{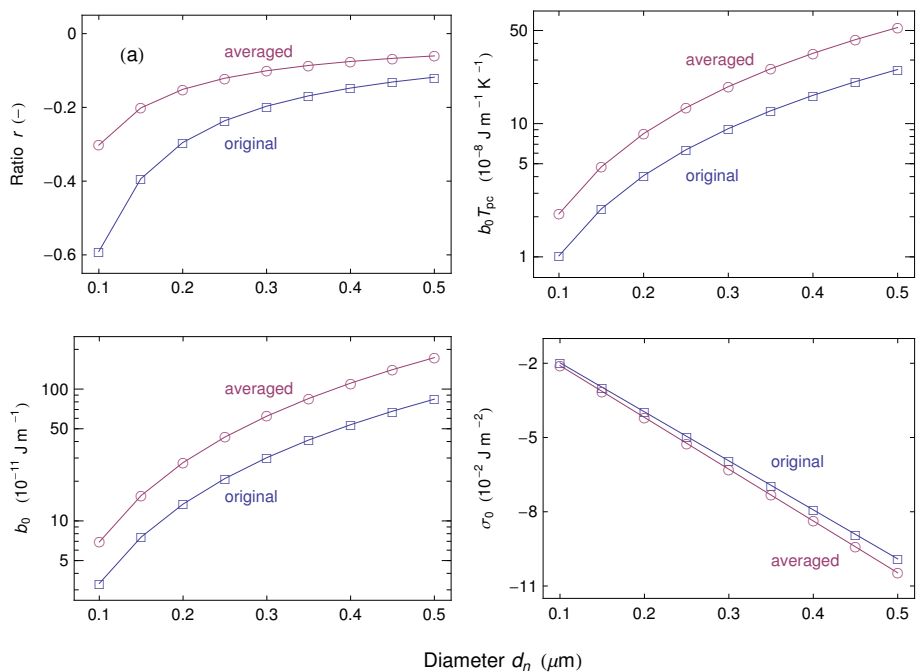}\\
\caption{(a) The ratio $r$ for various choices of $d_n$ as obtained by fitting the theoretical dependence $\al(r)$ from Fig.~\ref{fig: al}(b) to the experimental value of $\al$. (b) The product $b_0 T_\pc$ calculated from fitting the height of the peak $H$ to its experimental value. (c) The width $b_0$ calculated from $b_0 T_\pc$ and $T_\pc$. (d) The mean value $\sig_0$ obtained from  (The results for the original/averaged data are depicted as squares/circles.)}
  \label{fig: FitProc2}
\end{figure}

Third, we consider the \emph{asymmetry} factor, $0 < \al < 1$, of the peak in $c_\ex$. It is introduced as the ratio of the area under the peak that lies below the maximum position $T^*$ to the peak's total area $\ell$. Its value for the data in Fig.~\ref{fig: data norm}(d) and their averaged version in Fig.~\ref{fig: data norm av}(d) is $\al = 0.884$ and $\al = 0.821$, respectively. A theoretical expression for $\al$ follows from Eq.~\eqref{eq: c_exc explicit},
\begin{equation} \label{eq: al}
  \al \equiv \frac1\ell \int_0^{T^*} c_\ex (T) \, dT
  \approx \frac1{2 d_0^3} \sum_{i=1}^n d_i^3 \Big( \erf \frac{z_i}{\sqrt2} + 1 \Big),
\end{equation}
where $\erf$ is the Gauss error function. For the already obtained dependence $d^*(r)$ we now calculate the theoretical dependence of $\al$ on the ratio $r$. It is plotted in Fig.~\ref{fig: al}. Fitting these theoretical results to the experimental value of $\al$, we get the ratio $r$ for each diameter $d_n$, as is shown in Fig.~\ref{fig: FitProc2}(a). Note that the ratio $r$ is negative. This is when the grain peaks of which $c_\ex$ is a sum are shifted below the phase-change temperature, leading to $c_\ex$ with most of its area lying below its maximum position ($\al > 1/2$). A positive ratio $r$ would correspond to $c_\ex$ with most of its area lying above its maximum position ($\al < 1/2$).

Fourth, we consider the \emph{height} of the excess heat capacity peak for which Eq.~\eqref{eq: c_exc explicit} yields
\begin{equation} \label{eq: cP*}
  H = c_\ex (T^*) \approx
  \frac{\ell^2 \rho}{6\sqrt{2\pi} \; b_0 d_0^3 T_\pc} \,  \sum_{i=1}^n d_i^5 \, e^{- z_i^2/2}.
\end{equation}
Using the already determined ratio $r$ and dependence $d^*(r)$, this formula yields the peak height in the form $H = \const / b_0 T_\pc$ for each $d_n$. The experimental value $H = \SI{144.40}{kJ.kg^{-1}.K^{-1}}$ and  $H = \SI{120.54}{kJ.kg^{-1}.K^{-1}}$ taken from the data in Fig.~\ref{fig: data norm}(d) and averaged data in Fig.~\ref{fig: data norm av}(d), respectively, yield the product $b_0 T_\pc$ plotted in Fig.~\ref{fig: FitProc2}(b).

We may now use the experimental value of the maximum positions $T^* = \SI{300.4}{K}$ and $T^* = \SI{300.3}{K}$ for the original and averaged data to calculate the phase-change temperature from the expression $T_\pc = T^* - 6r(b_0 T_\pc) / \ell \rho d^*$ (see Eq.~\eqref{eq: T*}) and the already determined parameters. This yields practically the same values for all chosen diameters $d_n$ (the differences between $T_\pc$ for various $d_n$ do not exceed \SI{2}{mK}),
\begin{equation}
  T_\pc = \SI{303.7}{K} \; (\SI{30.5}{\degreeCelsius})
  \qquad \text{and} \qquad
  T_\pc = \SI{303.8}{K} \; (\SI{30.7}{\degreeCelsius}),
\end{equation}
respectively. Note that these values of $T_\pc$ are higher than the phase-change temperature determined in \cite{Glor11} by more than $\SI{3}{K}$ (see Table~\ref{tab: RT data}).

The determined values of $r$, $b_0 T_\pc$, and $T_\pc$ yield the width $b_0 = (b_0 T_\pc)/T_\pc$ and mean value $\sig_0 = b_0 r$. They are plotted in Fig.~\ref{fig: FitProc2}(c) and (d).

Finally, we must verify the condition from Eq.~\eqref{eq: cond} to see whether our theoretical formulas can be actually applied. From the fitted values of $b_0$ and $T_\pc$ we conclude that the ratio $\De T_i / \De\tau_i$ has lowest value $157.6$ (for $d_n = \SI{0.1}{\micro m}$ and $d_i = d_1 = \SI{10}{nm}$) so that the condition is indeed satisfied.

\begin{figure}
  \centering
  \includegraphics[width=\linewidth]{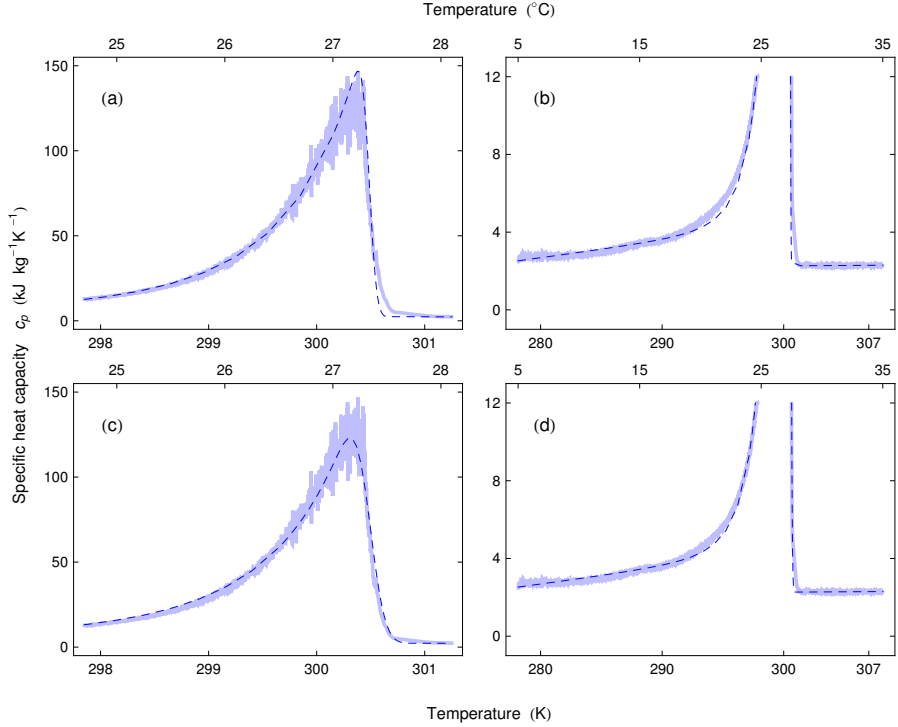}\\
\caption{The comparison of the experimental data on the specific heat capacity with theoretical results (the dashed lines) calculated from the fitted parameters obtained for the original data (parts (a) and (b)) and averaged data (parts (c) and (d)). The theoretical results are plotted for $d_n = \SI{0.1}{\micro m}$ (the other diameters yield practically the same peaks).}
  \label{fig: Fits cp}
\end{figure}

\begin{figure}
  \centering
  \includegraphics[width=\linewidth]{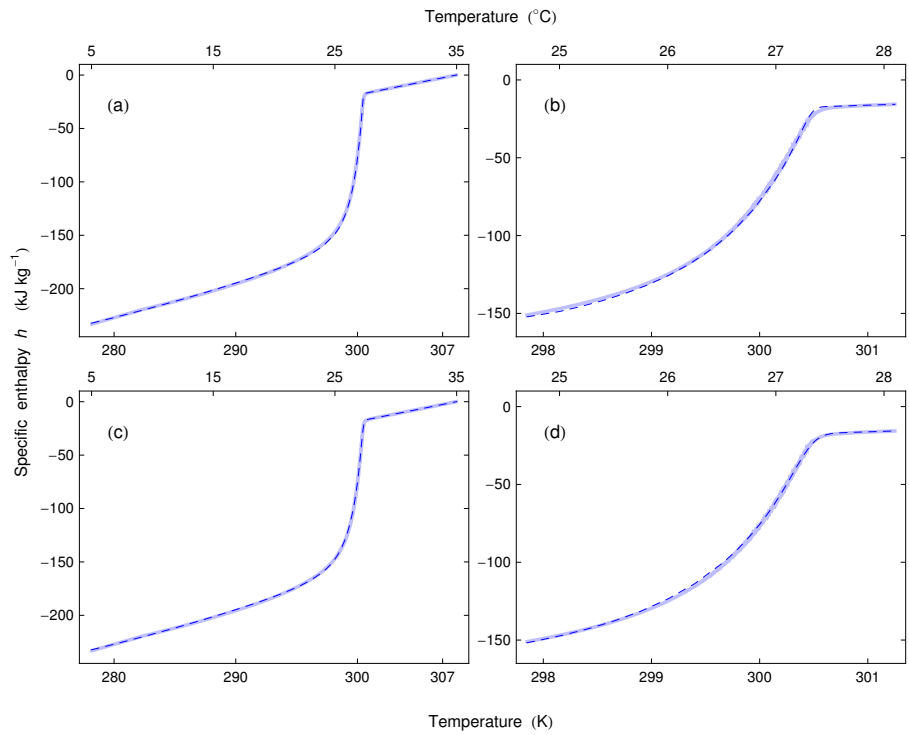}\\
\caption{The comparison of the experimental data on the specific enthalpy with theoretical results (the dashed lines) calculated from the fitted parameters obtained for the original data (parts (a) and (b)) and averaged data (parts (c) and (d)). The theoretical results are plotted for $d_n = \SI{0.1}{\micro m}$ (the other diameters yield practically the same peaks).}
  \label{fig: Fits h}
\end{figure}

Knowing the four parameters $T_\pc$, $\ell$, $\sig_0$, and $b_0$, we obtain the excess heat capacity $c_\ex$ from Eq.~\eqref{eq: c_exc explicit}, the jump function $J_\av$ from Eq.~\eqref{eq: cJ from c_exc}, and the heat capacity $c_p$ and enthalpy $h$ from Eq.~\eqref{eq: h,c poly}. The latter are plotted in Figs.~\ref{fig: Fits cp} and \ref{fig: Fits h}. They are practically identical for all chosen diameters $d_n$, so only the plots for $d_n = \SI{0.1}{\micro m}$ are shown. The agreement between the theoretical results and experimental data is very good.

\section{Conclusions}

We presented a quasistatic approach to describe the temperature dependence of the specific enthalpy $h$ and heat capacity $c_p$ of a parafin-based PCM Rubitherm~RT~27. We used experimental data for the heating run that were measured by adiabatic scanning calorimetry in which thermodynamic equilibrium of samples can be ensured. If the PCM were a single crystal, a microscopic theory of first-order phase transitions in finite systems would predict heat capacity spikes that are much sharper and taller (by several orders of magnitude) than those measured in experiments. Therefore, we used that the PCM should have a polycrystalline structure and modeled it as a large ensemble of small single-crystal grains. Then we were able to obtain theoretical results for $h$ and $c_p$ that could be fitted to experimental data with very good precision. We used only four fitting parameters, including the specific latent heat $\ell$ and phase-change temperature $T_\pc$. Their values were adjusted from four characteristics of the excess heat capacity peak (its area, maximum position, height, and asymmetry).

The key points of our approach may be summarized as follows.
\begin{enumerate}
\item
We provided a procedure to separate the baseline and excess heat capacities for a phase change between two phases, using the experimental data on the enthalpy and heat capacity.
\item
The presented equilibrium approach predicts an asymmetric jump and peak in the enthalpy and heat capacity, respectively, as a result of finite-size effects.
\item
The specific latent heat $\ell$ was identified with the area of the peak in the excess heat capacity. For the considered PCM it formed $84$ \% of the total enthalpy change in the range between $\SI{295.5}{K}$ and $\SI{305.5}{K}$.
\item
We determined the phase-change temperature $T_\pc$ from the height and maximum position of the peak in the excess heat capacity. Its value was higher by $\SI{3.2}{K}$ than the quoted one.
\end{enumerate}
Since the microscopic structure of the studied material was not taken into account in depth, our results are quite robust and could be applied to other PCMs. In addition, our results can be extended to the phase changes with coexistence of more than two phases, using the necessary modifications to the description of the single grain behavior. This may be a topic for a future investigation.

A weak point of our approach is the obtained value of the phase-change temperature that is rather shifted from the maximum position of the measured heat capacity peak. This is a consequence of taking the mean value $\sig_0$ to be fixed for all grain sizes. We may improve the results by considering $\sig_0$ to be varying with $L$ over a range of values. Then the half-width of this range would be an additional, fifth fitting parameter that must be determined from an additional property of the excess heat capacity peak. In this sense, the presented approach uses a minimal number of fitting parameters.

Another weak point is the tacit assumption that the PCM phases are crystalline, which may be a crude approximation, especially for the liquid phase. In addition, it is farfetched to apply the theory of phase changes from \cite{BoKo95} to the behavior of grains in a real PCM. Nevertheless, the precise shapes of the jump and peak functions $J$ and $P$ associated with the individual grains are not essential in the final results (see Eq.~\eqref{eq: c_exc explicit}), because such a detailed information is lost after their averages $J_\av$ and $P_\av$ over the grains are taken. Finally, the presented approach is restricted to the quasistatic regime. Thus, non-equilibrium effects are not considered, even though they may have an additional effect on the shape and position of the enthalpy jumps and heat capacity peaks.

\section*{Acknowledgements}

The research in this paper was supported by the Czech Science Foundation, Project No.~P105/12/G059, and by the VEGA project No.~1/0162/15. The authors would like to thank Prof.~Christ Glorieux and Dr.~Jan Leys from the Catholic University of Leuven, Belgium, for providing experimental data.



\end{document}